\documentclass[12pt]{article}
\begin{document}

\begin{center}
{\Large
Spin relaxation in n-doped gallium arsenide due to impurity
and electron-electron Elliot-Yafet scattering}

\bigskip
P.\ I.\ Tamborenea,$^{1,2}$ M.\ A.\ Kuroda,$^1$ and F.\ L.\ Bottesi$^1$ \\
$^1$Department of Physics ``J.\ J.\ Giambiagi'', University of
Buenos Aires \\
Ciudad Universitaria, Pab.\ I, C1428EHA Buenos Aires, Argentina \\
$^2$ Department of Physics and Astronomy, Clippinger Research Laboratories,
Ohio University, Athens, Ohio 45701-2979
\end{center}

\bigskip\noindent
{\it We calculate the spin relaxation time of conduction electrons
in $n$-doped bulk gallium arsenide.
We consider the Elliot-Yafet spin-relaxation mechanism, driven by
Coulombic-impurity and electron-electron scattering.
We find that these two scattering mechanisms result in relaxation
times of equal order of magnitude, but with disimilar dependences
on doping density and temperature.
Our theoretical results are compared with experimentally measured
spin relaxation times in gallium arsenide.

}

\bigskip
There is currently a great interest in the properties of
semiconductors derived from the electronic spin.
This interest stems largely from the foreseen growth of the
scope of applications of the spin degree of freedom in
electronics (spintronics) and computer science (quantum
computing) \cite{pri,aws-fla-sam}.
Among the properties of interest in relation to potential
applications, a central place is occupied by the spin relaxation
time, i.e., the characteristic time a spin-density imbalance
lasts inside a given material or structure.
Clearly, long relaxation times are generally desirable,
and an important progress in the search of systems with long
relaxation times has been made in a series of recent
experiments \cite{kik-smo-sam-aws,kik-aws-98,kik-aws-99}.
In those experiments it was found, among other things, that
the spin-relaxation time in semiconductors can be extended
by more than two orders of magnitude with appropriate
negative-type doping.

Several authors have recently analyzed this issue from a
theoretical point of view. Song and Kim \cite{son-kim} reviewed
the known spin-relaxation mechanisms and constructed a phase
diagram to depict graphically which mechanism is dominant in each
portion of parameter space. According to those authors, the
D'yakonov-Perel (DP) \cite{dya-per} mechanism dominates over the
Elliot-Yafet (EY) one \cite{ell,yaf} (driven by electron-impurity
scattering) except at low temperature in n-doped zincblende
semiconductors. Wu and Ning \cite{wu-nin} studied D'yakonov-Perel
spin relaxation in n-doped GaAs. Their main finding, as far as
explaining the above-mentioned experiments is concerned, is that
the DP mechanism produces and increasing relaxation time as a
function of applied magnetic field, in qualitative agreement with
experiment at high density \cite{kik-aws-98}. A related study,
based on a kinetic equation formalism and including the
k-dependence on the conduction-band g-factor, was done by Bronold
{\it et al.} \cite{bro-mar-sax-smi}. These authors find a good
agreement between their theory and experiment \cite{kik-aws-98}
for the relaxation time as a function of magnetic field at high
doping density. While the high-doping-density case can be treated
in terms of delocalized electrons---electrons at the bottom of the
conduction band---, the low-density regime calls for a description
based on electrons localized at the impurity sites. Kavokin
\cite{kav} and Gor'kov and Krotkov \cite{gor-kro} studied this
regime. The latter authors conclude that the Dzyaloshinskii-Moriya
interaction between localized electrons does not explain the
experimentally observed spin relaxation times, as was claimed by
Kavokin.

Here we report our calculations of the spin relaxation time of
electrons in the conduction band of n-type semiconductors caused by
the Elliot-Yafet (EY) scattering mechanism \cite{ell,yaf}
mediated by both electron-impurity and electron-electron
interactions.
Our calculations extend those of Chazalviel \cite{cha} and
Boguslawski \cite{bog} for electron-impurity and electron-electron
scattering, respectively.
Both of these authors considered only the case of almost
equal up-spin and down-spin populations.
Furthermore, Chazalviel studied only the zero temperature limit
while Boguslawski treated the high temperature case by introducing
Boltzmann rather than Fermi distributions.
We generalize those calculations by considering arbitrary
spin populations described by finite temperature Fermi
distributions.
Also, while those authors were mainly interested in indium
antimonide, we concentrate here on gallium arsenide, which
is one of the preferred semiconductor alloys in recent studies
of spin relaxation.

As it is clear that only an incomplete understanding of the
experiments reported in Ref.\ \cite{kik-aws-98} has emerged so
far, our aim here is to provide a missing piece in the set of
theoretical scenarios considered up to now. We point out that none
of the recent theoretical studies of spin relaxation in n-type
GaAs has taken into account the {\em electron-electron} EY
mechanism, which had been studied by Boguslawski \cite{bog} in a
different context.

The Coulomb interaction is independent of spin and therefore
cannot cause spin-flip transitions between conduction-band states
that are spin eigenstates.
The possibility of a spin flip in the EY scattering
mechanism arises from the fact that the conduction-band
states of some semiconductors are not spin eigenstates, which in
turn is due to the spin-orbit contribution to the crystal
Hamiltonian.
The conduction band states of zincblende semiconductors are
linear combinations of
both spin eigenstates with coefficients that are functions of
the crystal momentum ${\mathbf k}$.
Usually one of the two components in these admixtures has a much
larger amplitude than the other.
Therefore the mixed states retain a clear correspondence to the
original pure-spin states, and are still referred to as the
``spin-up" and ``spin-down" states.
We denote these states by $|{\mathbf k}\pm\rangle$.
The matrix element of a spin-independent scattering potential
(like that of ionized impurities and electron-electron interaction)
between these states can thus be non-zero, thereby activating
the spin-flip EY scattering mechanism.

Our general goal is to understand the spin relaxation process
in $n$-doped bulk semiconductors studied recently in ultrafast
Faraday rotation experiments \cite{kik-aws-98}.
In this article we do not investigate the initial stages of the
relaxation process, when electron-hole scattering is predominant,
but concentrate rather on the slower decay that takes place after
the electron-hole recombination has been completed.
Therefore, the starting point of our calculation is the assumption
that there is a doped unpolarized electron gas of density $n$
present in the host semiconductor, and an additional density
$n_{ex}$ of photoexcited electrons occupying conduction-band
states of the type $|{\mathbf k}+\rangle$.
Since we ``look" at the system at least several hundreds of
picoseconds after the pump pulse has created the photoexcited
polarized electrons, we assume that the two spin species have
reached thermal equilibrium and can be described by Fermi-Dirac
distributions
$f_{\pm}({\mathbf k})$ with a common temperature $T$.
We restrict our treatment to cases where $n_{ex} < n$.

We are interested in the relaxation of a spin-density difference
$n_d \equiv n_+ - n_-$,
where $n_{\pm}=(1/V)\,\sum_{\mathbf k}f_{\pm}(k)$ are the
densities of spin-up and spin-down electrons in the conduction
band.
We assume that initially $n_d = n_{ex}$.
The relaxation rate is defined by
\begin{equation}
\frac{1}{T_1} \equiv - \frac{\dot{n}_d}{n_d}
          =  - \frac{\dot{n}_+ - \dot{n}_-}{n_d}
          =  - \frac{2 \dot{n}_+}{n_d}.
\label{eq:T1_def}
\end{equation}
Here we have used that $\dot{n}_+ + \dot{n}_-$=0 since no further
excitation nor recombination take place in the considered
time regime.

For electron-impurity scattering, the relaxation rate is calculated as
\begin{eqnarray}
\frac{1}{T_1} &=& -\frac{2 n_i}{n_d} \sum_{\mathbf k} \sum_{\mathbf k'}
     f_-(k') [1-f_+(k)] \; T_{{\mathbf k'}- \rightarrow {\mathbf k}+}
    - f_+(k)  [1-f_-(k')] \; T_{{\mathbf k}+ \rightarrow {\mathbf k'}-}
            \nonumber \\
&=& \frac{2 n_i}{n_d} \sum_{\mathbf k} \sum_{\mathbf k'}
     [f_+(k)-f_-(k')] \;
     T_{{\mathbf k}+ \rightarrow {\mathbf k'}-},
\end{eqnarray}
where we have denoted the density of impurity scatterers by $n_i$,
and used that
$T_{{\mathbf k'}- \rightarrow {\mathbf k}+} = T_{{\mathbf k}+
\rightarrow {\mathbf k'}-}$.
The spin-flip transition rate due to the impurity EY mechanism is
\begin{equation}
T_{{\mathbf k}+ \rightarrow {\mathbf k'}-} =
    \frac{2\pi}{\hbar} \;
    \delta(E_{\mathbf k'}-E_{\mathbf k}) \;
    |\langle {\mathbf k'}-|V_i|{\mathbf k}+ \rangle|^2
\approx
       \frac{2\pi}{\hbar} \;
    \delta(E_{\mathbf k'}-E_{\mathbf k}) \;
    V_i({\mathbf k} - {\mathbf k'})^2 \;
    |\langle {\mathbf k'}-|{\mathbf k}+ \rangle|^2
\label{eq:sftp}
\end{equation}
The electron-impurity scattering is caused by the Coulomb potential
of the ionized silicon donors \cite{footnote1}, whose Fourier transform
is given by
$V_i({\mathbf q})= 4\pi e^2 / \epsilon \, V \, (q^2+k_s^2)$,
where $\epsilon$ is the lattice dielectric constant, $V$ is the volume,
and the screening wave vector in the quantum Debye-H\"uckel theory
for a degenerate electron gas is
$k_s = \left( 6 \pi \, e^2 \, n / \epsilon \, E_F \right)^{1/2}$.
For $n$ and $E_F$ in this expression we take the values corresponding
to the doped unpolarized electron gas.
The spin-mixed conduction-band states of zincblende semiconductors
can be calculated with the ${\mathbf k} \cdot {\mathbf p}$ perturbation
theory \cite{states}.
The scalar product of opposite-spin states is given by \cite{cha,bog}
\begin{equation}
\langle {\mathbf k'}-|{\mathbf k}+ \rangle =
                     \frac{\gamma \, \hbar^2}{4 m^{\ast} E_G} \;
                     (k_z {k'}_+ - {k'}_z k_+),
\label{eq:scalar_product}
\end{equation}
where
$\gamma = 2 \Delta (\Delta+2E_G) / (\Delta+E_G)(2\Delta+3E_G),$
$k_+=k_x+ik_y$,
$m^{\ast}$ is the conduction-band effective mass,
$\Delta$ is the valence-band spin-orbit splitting,
and $E_g$ is the bandgap energy.
Substituting Eq.~(\ref{eq:scalar_product}) into
Eq.~(\ref{eq:sftp}) yields
\begin{equation}
T_{{\mathbf k}+ \rightarrow {\mathbf k'}-} =
\frac{2\pi}{\hbar} \;
    \delta(E_{\mathbf k'}-E_{\mathbf k}) \;
    V_i({\mathbf k - \mathbf k'})^2 \;
    \left( \frac{\gamma \, \hbar^2}{4 m^{\ast} E_G} \right)^2 \;
    (k^2 {k'_z}^2 + k^{'2} k_z^2 - 2  k_z k'_z {{\mathbf k} \cdot {\mathbf k'}}).
\end{equation}
Doing the integral over $\mathbf k'$ and the angular part of the
integral over $\mathbf k$ one obtains
\begin{eqnarray}
\frac{1}{T_1} &=& \frac{n_i}{n_{d}} \frac{\hbar \gamma^2}{6\pi
m^{\ast} E_G^2}
           \frac{e^4}{\epsilon^2}
\int dk k^3 [f_+(k) - f_-(k)]
\left[-2 + (1+a) \, \ln{ \left(\frac{2+a}{a}\right) }
\right],
\label{eq:average_T1}
\end{eqnarray}
where $a \equiv (k_s/k)^2/2$.

For electron-electron scattering, the relaxation rate is calculated from
\begin{eqnarray}
 \frac{dn_+}{dt}=
\sum_{\stackrel{\mathbf{k}\,\mathbf{k'}}{\mathbf{q} \,
\sigma}}\!
\left\{
f_-(\mathbf{k}) f_\sigma(\mathbf{k'})
\left[1-f_+(\mathbf{k}+\mathbf{q})\right]
\left[1-f_\sigma(\mathbf{k'}-\mathbf{q})\right]
T_{\mathbf{k}\:-;\;\mathbf{k'}\sigma\rightarrow
\,\mathbf{k}+\mathbf{q}\:+;\;\mathbf{k'}-\mathbf{q}\;\sigma}-\right.
\nonumber\qquad\\ \qquad
\left.-\:f_+(\mathbf{k})f_\sigma(\mathbf{k'})\left[1-f_-(\mathbf{k}+\mathbf{q})
\right]
\left[1-f_\sigma(\mathbf{k'}-\mathbf{q})\right]T_{\mathbf{k}\:+;\;\mathbf{k'}\sigma
\rightarrow\mathbf{k}+\mathbf{q}\:-;\mathbf{k'}-\mathbf{q}\;\sigma}\right\},\qquad
\label{eq:eeEY}
\end{eqnarray}
\noindent
where
$\,T_{\mathbf{k_1}\;\pm\:;\:\mathbf{k'_1}\sigma\rightarrow\mathbf{k_2}\;\mp\:;
\,\mathbf{k'_2}\;\sigma}$
are the transition probabilities of scattering events where one of
the electrons flips spin and the other does not.
The likelyhood of double spin flips can be neglected.
Taking into account the direct and exchange terms, the transition
probability according to Fermi's golden rule is given by
\begin{eqnarray}
T_{\mathbf{k},-\:;\:\mathbf{k'},\sigma\rightarrow\:\mathbf{k}+\mathbf{q},+\:;
\,\mathbf{k'}-\mathbf{q},\sigma}=\frac{2\pi}{\hbar}\delta(E_f\!-\!E_i)\cdot
\qquad\qquad\qquad\qquad\qquad\qquad\qquad\qquad\nonumber\\
\cdot\left|\langle\mathbf{k}\!+\!\mathbf{q},+;
\mathbf{k'}\!-\!\mathbf{q},\sigma|\,V\,|\mathbf{k},-;\mathbf{k'},\sigma\rangle
-\langle\mathbf{k'}\!-\!\mathbf{q},
\sigma;\mathbf{k}\!+\!\mathbf{q},+|\,V\,|\mathbf{k},-;
\mathbf{k'},\sigma\rangle\right|^{\,2}.\label{fermirule1}
\end{eqnarray}
The interaction potential here is again the screened Coulomb potential
given above.
Its matrix elements between two-particle states are given again
by Eq.~(\ref{eq:scalar_product})
after a few standard approximations are made \cite{bog}.
The electron-electron EY relaxation time is obtained by solving
Eq.~(\ref{eq:eeEY}) numerically \cite{kur-tam}.

We first present results for the impurity EY mechanism
obtained numerically from Eq.~(\ref{eq:average_T1}).
That equation is valid for all zincblende semiconductor alloys,
but in this article we concentrate on gallium arsenide, given
the relative importance of this material in recent experimental
studies.
For a given material, the impurity EY spin relaxation time
depends on the dopant density $n_i$,
the density of the excess electrons with spin up, $n_d=n_{ex}$,
and the temperature $T$.
The density of donated electrons, $n$, which also enters implicitly
in the calculation, is taken to be $n=n_i$.
We envision that the initial spin-density difference, $n_d$, comes
from the photo-excitation caused by a circularly-polarized pump laser
pulse in a pump-and-probe experiment.

Our calculations show that, for impurity EY scattering, the
dominant factors in the determination
of $T_1$ (according to (\ref{eq:average_T1})) are the densities $n$
and $n_i$.
Therefore, we show first the time $T_1$ versus $n$, for three
different values of $n_d$ and of $T$ (Fig.~(1)).
At low temperature and low spin-density imbalance, $T_1$ decreases
by a factor of 1000 for a variation of a factor of 100 in $n$;
thus, $T_1$ follows roughly a power law on $n$ with an exponent
of -3/2.
This dependence comes from both $n$ and $n_i$.
The density of scattering centers, $n_i$, enters linearly in
Eq.~(\ref{eq:average_T1}), while the density of donated electrons
$n$ affects the integral in that equation through the screening
length $1/k_s$ and the Fermi functions.

The dependence of $T_1$ on $T$ is strongest at low values of $n$,
as could be expected from reasoning on the effect of
temperature on Fermi functions for different densities.
For $n=10^{18} \, \mbox{cm}^{-3}$ there is practically no temperature
dependence up to the highest temperature considered here, $T=$100~K.
The relaxation time also depends very weakly on $n_d$, except
when $n_d$ approaches $n$.

In Fig.~2 we present more detailed data on $T_1$ versus the
spin-density imbalance, $n_d$, for doping density $n=10^{16} \mbox{cm}^{-3}$.
Temperatures from 1~K to 100~K are considered, as indicated
in the figure caption.
The main conclusion we can draw from Fig.~2 is that there is
essentially no variation of $T_1$ with $n_d$ for $n_d < n/10$.
(This conclusion is also valid for $n=10^{17}, 10^{18} \mbox{cm}^{-3}$,
data not shown.)
The variation of $n$ between $n_d= n/10$ and $n_d= n$ is roughly
of a factor 2 at low temperature, and almost none at high temperature.
(The definition of low and high $T$ depends on $n$, as mentioned
above.)

We shall now move on to the results for the electron-electron EY scattering.
We define the relative density $n_{exr}$ as the ratio between the
density of photoexcited electrons and the doping density,
$n_{exr} \equiv n_{ex}/n$.
Figure 3 shows the spin relaxation time $T_1$, obtained by solving
Eq.~(\ref{eq:eeEY}) and using Eq.~(\ref{eq:T1_def}), as a function of $n$ for
$n_{exr}=0.001, 0.01, 0.1$ and various temperatures between 1~K and 315~K.
Once again we obtain that $T_1$ decreases rapidly with doping density
much as it does in the case of impurity EY scattering.
On the other hand, Fig.~3 shows also a strong temperature dependence,
not present for impurity scattering.

In Fig.~3 we can identify different regimes that can be broadly classified
according to the doping density and the temperature.
A high temperature regime appears for
$T > 10 \, K$ and $n \,^<_\sim \, 10^{16}\mbox{cm}^{-3}$,
or $T > T_F$, where $T_F$ is the Fermi energy of the doped electron gas.
The low temperature regime is located on the ``opposite corner (upper-right)''
of the plot, and is given roughly by
$T < 3.2K$ and $n \,^>_\sim \, 10^{17} \mbox{cm}^{-3}$ ($T < T_F$).
Other combinations of $n$ and $T$ correspond to $T \approx T_F$.
The following main conclusions can be drawn from Fig.~3:
(i) At high temperature ($T_F < T$), $T_1$ does not depend significantly
on $n_{exr}$;
(ii) At low temperature ($T < T_F$), as $n$ increases $T_1$ decreases
and this decrease is sensibly greater the higher the value of $n_{exr}$.
(iii) At low temperature (see curves for $T=1~K, 3.16~K$),
high $n_{exr}$ (see $n_{exr}=0.1$),
and high density ($n \approx 10^{18} \mbox{cm}^{-3}$, the temperature
dependence of $T_1$ dissapears.
We have analyzed in detail the dependence of $T_1$ on $n$, $T$, and $n_{exr}$,
but at this point we shall just refer the interested reader
to Ref.~\cite{kur-tam}.

Finally, we compare our theoretical results with the corresponding
experimental findings of Ref.~\cite{kik-aws-98}. The longest
spin-relaxation times reported in Ref.~\cite{kik-aws-98} where
obtained at low magnetic field and with a doping density of $n =
10^{16} \mbox{cm}^{-3}$. In Fig.~4 we plot the relaxation time
versus temperature. The two solid curves are results from our
electron-electron EY scattering calculation and the triangles are
experimental data from Ref.~\cite{kik-aws-98}. There is a large
disagreement between the experimental points and the calculated
curves for all temperatures, of two to three orders of magnitude.
We recall that the impurity-EY relaxation times (Figs.~1 and 2)
are of the same order of magnitude as the electron-electron EY
relaxation times shown in Fig.~4, and therefore they also compare
unfavorably with experiment. The fact that both types of EY
relaxation times are much longer than the ones measured
experimentally could indicate that another, more efficient,
spin-flip mechanism dominates the spin relaxation in this system.
A partial agreement between experiment and theory has been found
in terms of the D'yakonov-Perel spin-flip mechanism in a
calculation that includes a careful treatment of the semiconductor
band structure and experimentally measured mobilities as input
parameters \cite{lau-ole-fla}. That treatment seems to fit the
experimental relaxation times at a temperature of around $100 K$,
but is off by two orders of magnitude at low temperature.
Therefore, we cannot rule out yet that a refined version of the EY
calculation presented here could help explain the measured
relaxation times in n-doped GaAs, particularly at low temperature.
The refinement in the calculation should probably be made on the
expression of the spin-mixed conduction-band states, the treatment
of screening of the Coulomb potential, and also the Born
approximation to scattering rates used here and in
Refs.~\cite{cha,bog}. Furthermore, at low density the possibility
of electron localization at impurity sites has to be taken into
account \cite{kav,gor-kro}.

In Ref.\ \cite{kik-aws-98} it is mentioned that the
electron-electron EY mechanism could help understanding the
temperature dependence of the relaxation time at low temperature
(below 30~K). However, as noted by the authors, the estimates
based on the treatment of Boguslawski \cite{bog}, which assumes
classical statistics, yield a relaxation rate that is actually too
fast compared to experiment. Our calculation of the
electron-electron EY mechanism shows that Boguslawski's results
cannot be used to interpret experiments at low temperatures ($<100
K$). Indeed, the use of Fermi distributions enforces Pauli
blocking which greatly reduces the efficiency of electron-electron
scattering as compared to the case of Boltzmann statistics.

In conclusion, we have generalized the treatments of Chazalviel
\cite{cha} and Boguslawski \cite{bog} of impurity and
electron-electron EY spin-flip scattering in n-doped zincblende
semiconductors. Our treatment of spin relaxation takes into
account the dependence of $T_1$ on doping density, spin-density
imbalance, and temperature. We apply this theory to GaAs and
analyze the main features of the dependence of $T_1$ on those
system variables. This theory should mainly be considered as a
starting point for more refined treatments of the Elliot-Yafet
mechanism, as a complete understanding of recent experiments
\cite{kik-smo-sam-aws,kik-aws-98} based on the known spin-flip
mechanisms in semiconductors has not emerged so far.

\bigskip\noindent
{\bf Acknowledgements}

\medskip
We thank Jay Kikkawa and Horia Metiu for useful discussions. 
PIT thanks QUEST, the NSF Center for Quantized Electronic 
Structures at UCSB, and Professor Metiu for support during 
the initial stages of this work. 
Partial support from Fundaci\'on Antorchas, Proyectos
UBACyT 2001-2003, and CONICET is also acknowledged. 
PIT also thanks Sergio Ulloa for support at Ohio University.

\newpage
\noindent
FIGURE CAPTIONS

\bigskip\noindent
FIGURE 1. Impurity EY spin relaxation time of GaAs versus doping density at
three different temperatures ($5\,K$, $50\,K$, and $100\,K$)
and three spin-density differences:
$n_d=10^{14} \, \mbox{cm}^{-3}$ (solid lines),
$10^{15} \, \mbox{cm}^{-3}$ (dotted lines), and
$10^{16} \, \mbox{cm}^{-3}$ (dashed lines).

\bigskip\noindent
FIGURE 2. Impurity EY spin relaxation rate of GaAs versus spin-density
difference ($n_d=n_{ex}$) for doping density $n=10^{16} \, \mbox{cm}^{-3}$
(equal to the impurity density $n_i$.)
The various curves correspond, from top to bottom, to
$T=1, 10, 20, \ldots, 100\,\mbox{K}$.

\bigskip\noindent
FIGURE 3.  Electron-electron EY relaxation time versus doping density $n$
for several values of the temperature and the relative density of photoexcited
electrons, $n_{exr}$.

\bigskip\noindent
FIGURE 4. Spin relaxation time versus temperature.
Comparison between experimental data (Ref.~\cite{kik-aws-98})
and results of the electron-electron EY scattering calculation.
The theoretical curves are for
$n=10^{16} \, \mbox{cm}^{-3}$ and
$n_{ex}=10^{14}$ and $10^{15} \, \mbox{cm}^{-3}$.

\newpage

\begin{figure}
 \vbox to 8cm {\vss\hbox to 10cm
 {\hss\
   {\includegraphics{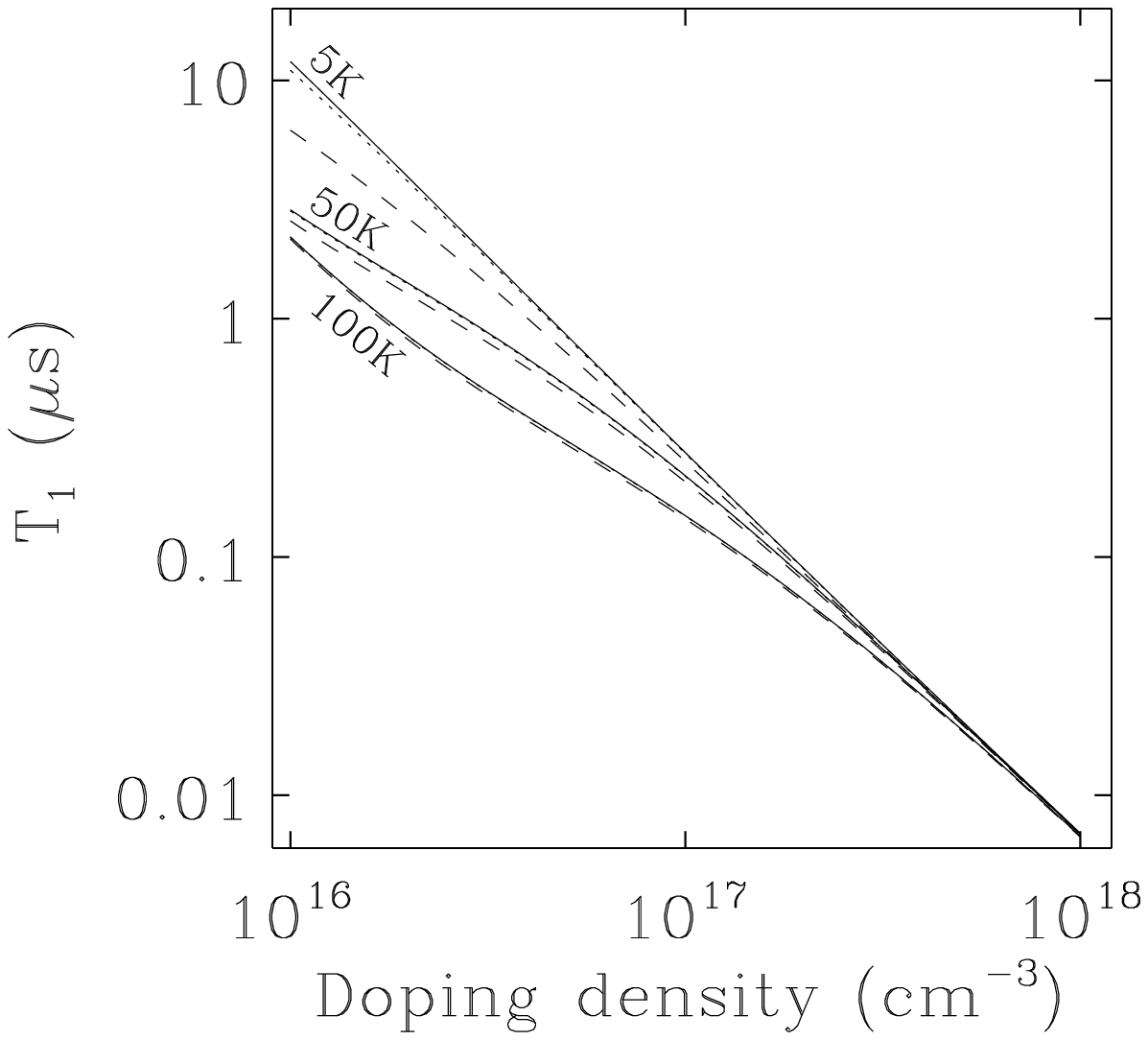}
   }
  \hss}
 }
\caption{}
\end{figure}
\begin{figure}
 \vbox to 8cm {\vss\hbox to 10cm
 {\hss\
   {\includegraphics{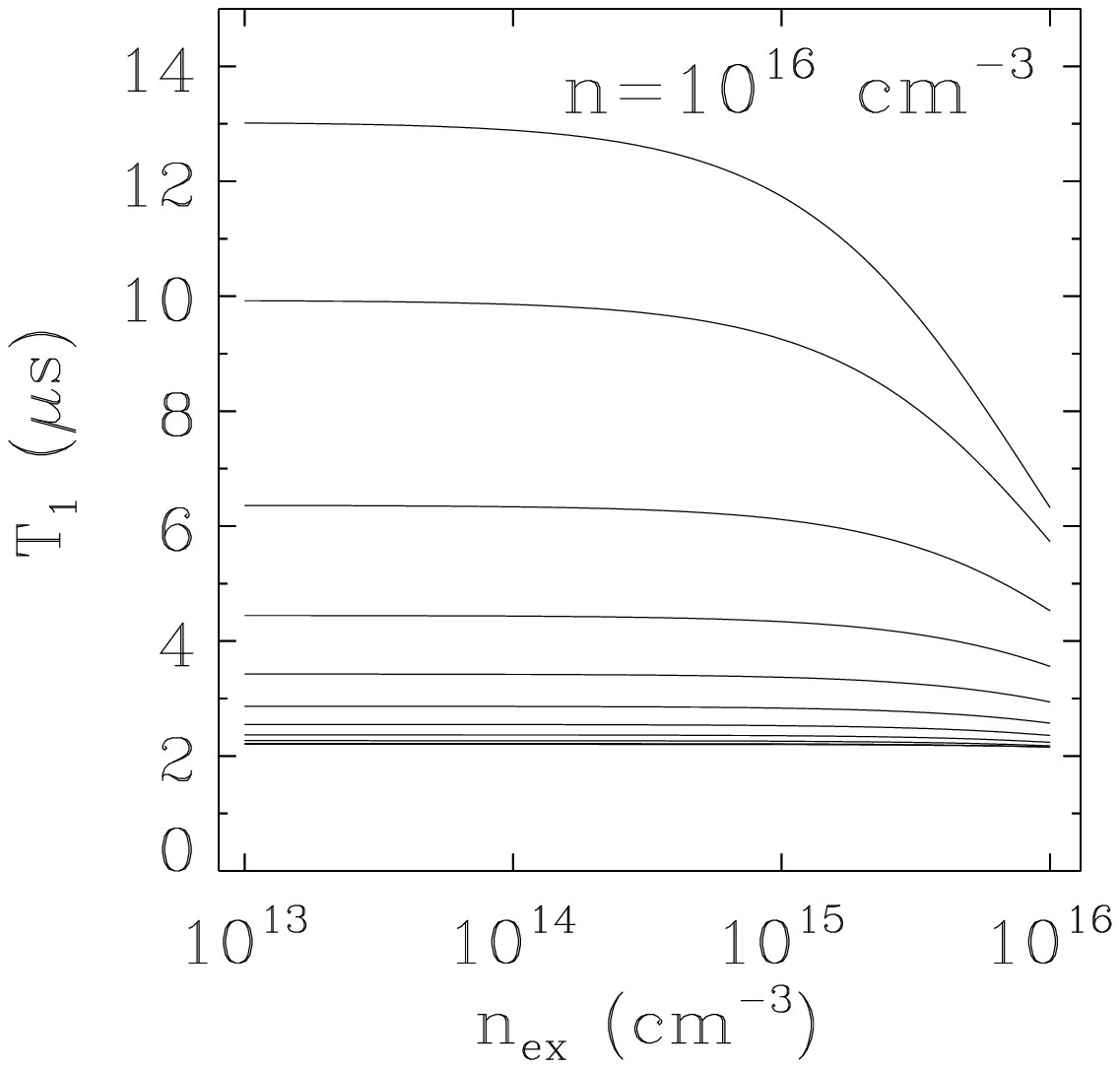}
   }
  \hss}
 }
\caption{}
\end{figure}
\begin{figure}
 \vbox to 10cm {\vss\hbox to 10cm
 {\hss\
   {\includegraphics{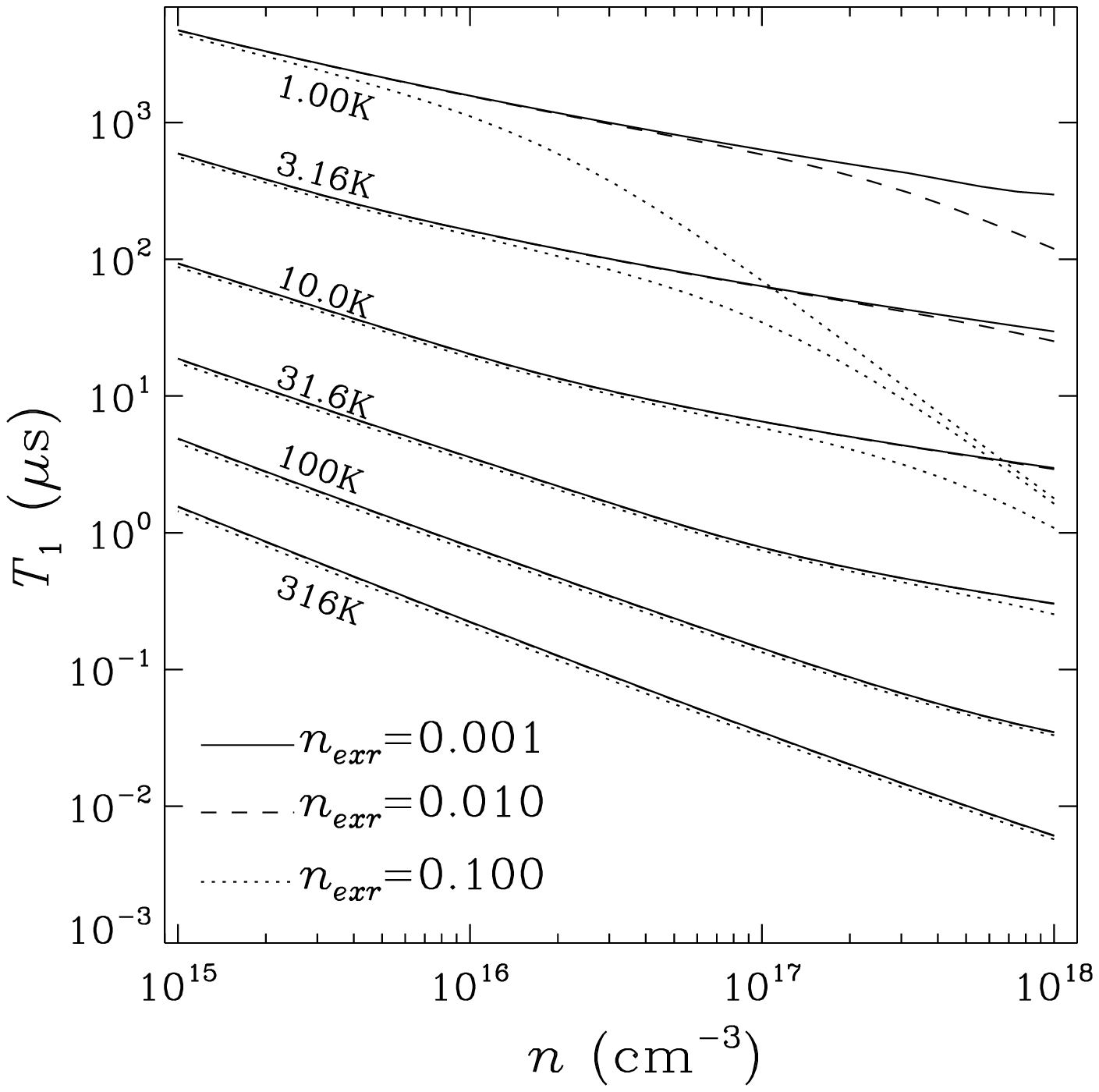}
   }
  \hss}
 }
\caption{}
\end{figure}
\begin{figure}
 \vbox to 10cm {\vss\hbox to 10cm
 {\hss\
   {\includegraphics{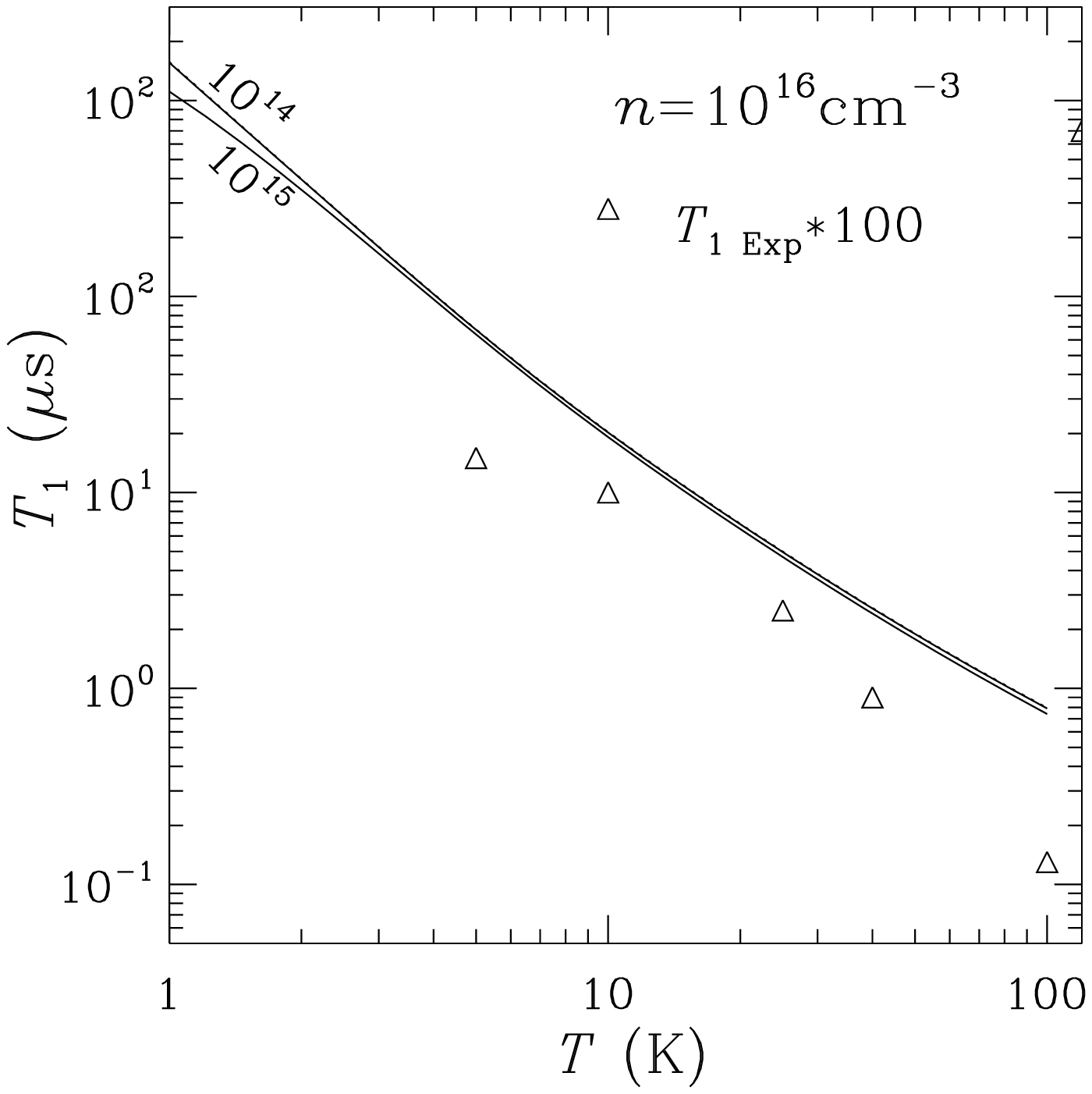}
   }
  \hss}
 }
\caption{}
\end{figure}

\end{document}